# Filtering for Copyright Enforcement in Europe after the Sabam cases

Stefan Kulk[a], Frederik Zuiderveen Borgesius[b]

[a]PhD Researcher, Centre for Intellectual Property Law (University of Utrecht)
s.kulk[at]uu.nl

[b]PhD Researcher, Institute for Information Law (University of Amsterdam)
f.j.zuiderveenborgesius[at]uva.nl

**Abstract.** Sabam, a Belgian collective rights management organisation, wanted an internet access provider and a social network site to install a filter system to enforce copyrights. In two recent judgments, the Court of Justice of the European Union decided that the social network site and the internet access provider cannot be required to install the filter system that Sabam asked for. Are these judgments good news for fundamental rights? This article argues that little is won for privacy and freedom of information.

## 1. Introduction

In the *Scarlet v Sabam*[1] and *Netlog v Sabam*[2] judgements, the Court of Justice of the European Union dealt with the question of whether an obligation to install a mechanism to filter out copyright protected works is in line with EU law. In these judgements, the Court struck through a general obligation to filter out copyrighted material. Some have greeted the judgments with enthusiasm.[3] But are these judgements indeed good news for privacy and information freedom? This article argues that little is won for these fundamental rights.

Section 2 provides background information to the cases, and section 3 places the cases in the context of European case law on the liability of internet intermediaries. Section 4 discusses fundamental rights. Section 5 contains a conclusion: the Court gave two reasonable decisions but could have made a more thorough analysis when it considered the fundamental rights of internet users. Therefore, little is won for privacy and freedom of information.

## 2. Background of the cases

The Court of Justice of the European Union decided two Belgian cases that dealt with the legality of filter obligations. In both cases, Sabam, a Belgian collective rights management organisation, asked the national court to require an intermediary to monitor the internet use of its users, and to stop and prevent the exchange of copyrighted works. In both cases, the national court asked preliminary questions to the Court of Justice of the European Union. In short, the central question in the cases was as follows. Do the

---

[1] *Scarlet v Sabam* (C-70/10) [2012] E.T.M.R. 4.

[2] *Sabam v Netlog* (C-360/10) [2012] 2 C.M.L.R. 18.

[3] For instance, European Digital Rights (EDRI) calls the Scarlet/Sabam judgement 'a vital victory for internet freedoms', http://www.edri.org/scarlet_sabam_win, accessed 26 June 2012.

Information Society Directive[4] and the Enforcement Directive[5], in the light of the E-Commerce Directive,[6] the Data Protection Directive,[7] the E-Privacy Directive,[8] and fundamental rights as protected in the European Convention on Human Rights, preclude such a filter obligation?

In *Scarlet*, Sabam wanted internet access provider Scarlet to install a filter system to prevent its users from sharing copyright protected music through peer-to-peer software. The European Court explained the filter system in four steps. (i) The internet access providers identifies, within all of the electronic communications of all its customers, the files relating to peer-to-peer traffic; (ii) it identifies, within that traffic, the files containing works of which holders of intellectual-property rights claim to hold rights, (iii) it determines which of those files are shared unlawfully, and (iv) it blocks file sharing that it considers to be unlawful.

*Scarlet* concerns a filter obligation imposed on an internet access provider, whereas *Netlog* concerns a similar obligation imposed on a social network provider. The judgement in *Netlog* resembles the earlier and more extensive judgement in *Scarlet*. This article only mentions *Netlog* where it adds something to *Scarlet*.

## 3. Intermediary Liability in the EU

The two judgments are part of a series of judgments in which the Court explains the responsibilities of internet intermediaries in the battle against the infringement of intellectual property rights.

The rules governing the liability of internet intermediaries for copyright infringement are a patchwork of European Directives, national implementing legislation and different national liability regimes. At the European level, mainly the E-commerce Directive is important as it limits the liability of certain types of internet intermediaries.[9] The Directive doesn't establish liability, but determines the circumstances in which the liability of intermediaries should be limited.[10] Intermediaries may benefit from the liability exemption when they act as a 'mere conduit', use 'caching' techniques or offer 'hosting' services.

The liability of internet access providers such as Scarlet is limited on the basis of art. 12 E-commerce Directive. Access providers aren't liable for the information they transmit if they (i) don't initiate the transmission; (ii) don't select the receiver of the transmission; and (iii) don't select or modify the transmitted information.

A provider of hosting services (storage of content) is not liable for content on the basis of art. 14 E-commerce Directive, if the following conditions are met. (i) It doesn't have knowledge that the content is illegal, (ii) it acts quickly to remove information when receiving a notice, and (iii), and the content isn't posted under its authority.[11]

---

[4] Directive 2001/29/EC on the harmonisation of certain aspects of copyright and related rights in the information society [2001] OJ L 167/10.

[5] Directive 2004/48/EC on the enforcement of intellectual property rights [2004] OJ L 157/45.

[6] Directive 2000/31/EC on certain legal aspects of information society services, in particular electronic commerce, in the Internal Market [2000] OJ L 178/1.

[7] Directive 95/46/EC on the protection of individuals with regard to the processing of personal data and on the free movement of such data [1995] OJ L 281/31.

[8] Directive 2002/58/EC concerning the processing of personal data and the protection of privacy in the electronic communications sector [2002] OJ L 201/37.

[9] Art. 12 to 14 E-Commerce Directive.

[10] For instance, the first sentence of art. 12 of the E-Commerce Directive reads: "where an information society service is provided that consists of the transmission in a communication network of information provided by a recipient of the service, or the provision of access to a communication network, Member States shall ensure that the service provider is not liable for the information transmitted."

[11] On 1 June 2012, the European Commission started a consultation about the hosting safe harbour: "A clean and open Internet: Public consultation on procedures for notifying and acting on illegal content hosted by online

*Louis Vuitton v Google France* was the first case in which the Court elaborated on the liability exemptions in the E-Commerce Directive.[12] It concerned the question whether Google could be considered a hosting provider, when it acts as a search engine advertising provider. A second question was whether Google could rely on the liability exemption for trademark infringement committed by its users.

Art. 14 E-Commerce Directive focuses on traditional hosting providers, who were among the main players on the internet at the time the directive was drafted. It was not evident that the provider of a search engine advertising service could rely on the hosting liability exemption, as it has more influence on what happens on its platform than traditional hosting providers. Nevertheless, the Court qualified Google as a hosting provider. But the Court also said that for limitation of liability, the provider should be neutral and should therefore take a mere technical, automatic and passive role.[13] The Court based this requirement on recital 42 of the E-Commerce Directive, which strictly speaking only applies to mere conduits and caching providers. The Court of Justice of the European Union left it for the national courts to decide whether Google's liability is actually limited.

In *L'Oréal v eBay* the Advocate General expressed his doubts about the neutrality requirement for the limitation of liability of hosting providers.[14] But the Court confirmed its earlier interpretation of art. 14. It also said that the liability exemption for hosting providers "must, in fact, be interpreted in the light not only of its wording but also of the context in which it occurs and the objectives pursued by the rules of which it is part."[15] The Court didn't explain which objectives and context it meant.

## 4. The Sabam Cases in the Context of the European Regime for Intermediary Liability

In *Scarlet*, the European Court is quick to establish that Scarlet that the rules of art. 12 to 15 E-Commerce Directive apply to the internet access provider.[16] In *Netlog*, one of the questions was whether a social network provider can be considered a hosting provider. The Court's answer is clear: "it is not in dispute that the owner of an online social networking platform (…) stores information provided by the users of that platform, relating to their profile, on its servers, and that it is thus a hosting service provider within the meaning of Article 14 (…)."[17] The most important lesson from *Netlog* is that providers of social network sites can rely on the hosting liability exemption.

Regardless of the liability exemptions in the E-Commerce Directive, and regardless of whether the intermediary is responsible for an infringement, national judges can impose injunctions on the intermediary. This follows from the E-Commerce Directive,[18] but also from the Information Society Directive[19] and the Enforcement Directive,[20] which harmonises civil enforcement of intellectual property rights.

---

intermediaries", http://ec.europa.eu/internal_market/consultations/2012/clean-and-open-internet_en.htm, accessed 6 June 2012.

[12] *Google France v Louis Vuitton* (C-236/08 to C-238/08) [2010] E.T.M.R. 30.

[13] *Google France* [2010] E.T.M.R. 30 at [113].

[14] *L'Oréal v eBay* (C-324/09) [2011] E.T.M.R. 52, opinion of AG Jääskinen paras 139-140.

[15] *L'Oréal v eBay* [2011] E.T.M.R. 52 at [111].

[16] *Scarlet v Sabam* [2012] E.T.M.R. 4 at [34].

[17] *Sabam v Netlog* [2012] 2 C.M.L.R. 18 at [27].

[18] Art. 12(3), 13(2), and 14(3) E-Commerce Directive.

[19] Art. 8 Information Society Directive.

[20] Art. 9 and 11 Enforcement Directive.

The Information Society Directive and the Enforcement Directive give rights holders the possibility to ask for an injunction against an infringer, aimed at prohibiting the continuation of an infringement.[21] Rights holders can also apply for an injunction against intermediaries whose services are used by a third party to infringe intellectual property rights.[22] Thus, while the liability of Scarlet and Netlog is limited, obligations to stop copyright infringement can still be imposed on these service providers. The filter obligations that Sabam requested are examples of such injunctions.

When an internet access provider's subscribers infringe copyright, does this mean that its service is 'used by a third party to infringe a copyright'? Should there be a close relationship between the intermediary and the third party that infringes? With regard to internet access providers, the Court dealt with this question in *LSG v Tele2*.[23] This case concerned an obligation to release the identities of Tele2 internet subscribers that were allegedly involved in peer to peer file-sharing. The Court's answer was clear. "Access providers which merely provide users with internet access, without offering other services such as email, FTP or file-sharing services or exercising any control, whether de iure or de facto, over the services which users make use of, must be regarded as 'intermediaries' within the meaning of Article 8(3) of the [Information Society] Directive."[24]

A filter obligation seeks to *prevent* rather than to *stop* copyright infringement. Nevertheless, the Court held in *L'Oréal v eBay* that national courts must allow rights holders to take measures that contribute to preventing further infringements. Such measures must, however, be "fair and equitable", not unnecessarily complicated or costly, nor entail unreasonable time limits.[25] The filter obligation that Sabam requested doesn't meet these requirements. According to the Court, the filter system is a complicated, costly, and permanent computer system. Moreover, Sabam wanted Scarlet to pay for the system.[26]

Injunctions must also meet requirements that are not derived from the Enforcement Directive. Art. 15 of the E-Commerce Directive prohibits imposing a general obligation to monitor on intermediaries.

In *L'Oréal v eBay* the Court said that "the measures required of the online service provider concerned cannot consist in an active monitoring of all the data of each of its customers in order to prevent any future infringement of intellectual property rights."[27] In *Scarlet* the Court concludes that the filter obligation that Sabam asked for would be contrary to art. 15(1) E-Commerce Directive because it entails monitoring *all* data from *all* customers for *any* future infringement of intellectual property, for an unlimited time.[28] In sum, the E-Commerce Directive limits the measures that can be taken under the Enforcement Directive.

It isn't surprising that the Court sees the broad filter obligation requested by Sabam as a general obligation to monitor. The judgment therefore provides little guidance when it comes to more specific filter obligations. A filter system that is specific with respect to the group of persons (suspected uploaders), but general in respect of the content (all internet traffic of that group), may still be possible after the judgment. Sabam's fate was sealed after the Court established that the filter obligation would be in breach of the E-Commerce Directive.

---

[21] Art. 8 Information Society Directive, and art. 11 Enforcement Directive.

[22] Art. 8(3) Information Society Directive, and art. 11 Enforcement Directive.

[23] *LSG v Tele2* (C-557/07) [2009] ECRI-1227.

[24] *LSG v Tele2* [2009] ECRI-1227 dictum.

[25] Art. 3(1) Enforcement Directive.

[26] *Scarlet v Sabam* [2012] E.T.M.R. 4 at [48].

[27] *L'Oréal v eBay* [2011] E.T.M.R. 52 at [139].

[28] *Scarlet v Sabam* [2012] E.T.M.R. 4 at [40].

## 5. Fundamental Rights

Although the Court focuses on the E-Commerce Directive, it briefly discusses fundamental rights as well. The Court relies mainly on the freedom to conduct a business, and on the right to intellectual property. The latter is enshrined in art. 17 paragraph 2 of the Charter of Fundamental Rights of the European Union: "Intellectual property is protected." But, the Court says that "nothing whatsoever in the wording of that provision or in the Court's case-law (…) suggest[s] that that right is inviolable and must for that reason be absolutely protected."[29] This conclusion is in line with the legislative history of the Charter.[30]

In *Promusicae*, the Court held that the right to property and thus the enforcement of intellectual property rights has to be balanced against other fundamental rights.[31] Furthermore, the directives must be interpreted in such a way that they are not in conflict with "general principles of Community law, such as the principle of proportionality."[32]

Art. 16 of the Charter contains a fundamental right called "freedom to conduct a business". The Court speaks of "a serious infringement of the freedom of the ISP concerned to conduct its business since it would require that ISP to install a complicated, costly, permanent computer system at its own expense."[33] Therefore the balance between the different fundamental rights would be lost if the filter obligation would be imposed.

The Court pays less attention to the fundamental rights of the subscribers of Scarlet. The Belgian court referred to the European Convention on Human Rights in its question, and the Attorney General discussed the Convention in his opinion. But the Court refers only to the Charter. The Court considers only the right to data protection of the subscribers, and their freedom to receive information (art. 8 and 11 of the Charter). These two rights are each addressed in one paragraph. The Court notes that the filter obligation "could potentially" limit freedom of information, because the filter may not adequately distinguish legal from illegal content, so its application could lead to the blocking of communications with legal content.[34] The filter system Sabam asked for is a drastic measure. Therefore, a more extensive analysis of the restriction of freedom of information would have been welcome.

Regarding data protection, the Court says that the "filtering system would involve a systematic analysis of all content and the collection and identification of users' IP addresses from which unlawful content on the network is sent."[35] Prompted by the Attorney General, the Court decides that the IP addresses are personal data, a decision that it had steered away from until now.[36] "Those addresses are protected personal data because they allow those users to be precisely identified."[37] The Advocate General relies mainly on Opinions of the Article 29 Working Party, the advisory and consultative body of European privacy regulators.[38] These Opinions, which are not legally binding, thus influence the Court. The Court doesn't explain why Scarlet cannot process personal data in this case. However, the Data Protection Directive doesn't prohibit the processing of personal data without consent. The Court doesn't finish its argument regarding data protection. It fails to explain why the Data Protection Directive doesn't provide a ground that could legitimise the processing of personal data by Scarlet.

---

[29] *Scarlet v Sabam* [2012] E.T.M.R. 4 at [43].

[30] Note from the Praesidium, Draft Charter of Fundamental Rights of the European Union, doc. no. CHARTE 4473/00, Brussels, 11 October 2000, 19-20.

[31] *Promusicae v Telefonica de Espana* (C-275/06) [2008] E.C.D.R. 10 at [66-70]. Also see *Bonnier Audio v Perfect Communication Sweden* (C-461/10) [2012].

[32] *Promusicae v Telefonica de Espana* [2008] E.C.D.R. 10 dictum.

[33] *Scarlet v Sabam* [2012] E.T.M.R. 4 at [48].

[34] *Scarlet v Sabam* [2012] E.T.M.R. 4 at [52].

[35] *Scarlet v Sabam* [2012] E.T.M.R. 4 at [51].

[36] HvJ EU 9 November 2010, nr. C 92/09 en C 93/09 (Volker und Markus Schecke GbR & Hartmut Eifert/Land Hesse), at [42].

[37] *Scarlet v Sabam* [2012] E.T.M.R. 4 at [51].

[38] *Scarlet v Sabam* [2012] E.T.M.R. 4, opinion Advocate General paras 74-78.

The discussion on the status of IP addresses as personal data may not be over yet, even though most national judges in Europe, and the Article 29 Working Party, are of the opinion that IP addresses should be regarded as personal data. In *Scarlet*, the IP addresses are held by an access provider. For parties that do not offer internet access, it is harder to tie an IP address to a name. If such parties do not use IP addresses to distinguish, or to "single out", an individual, they could argue that they do not process personal data.[39] The status of IP addresses as personal data is relevant for parties that harvest IP addresses of peer-to-peer users for copyright holders.

The Court's ruling reaffirms that information without a name may be personal data. It follows that nameless profiles must probably be regarded as personal data as well. Such profiles are used for behavioural targeting, the monitoring of online behaviour of internet users to target them with customised advertising.[40] Incidentally, in the proposal for a Regulation that will replace the Data Protection Directive, IP addresses are considered to be personal data in most cases.[41]

Back to the judgment. The Court's analysis doesn't go much deeper than the summary above. The Court does not discuss secrecy of communications or the right to respect for private life (art. 7 of the Charter). The European Court of Human Rights tends to be critical of systems to intercept communications,[42] also when such systems monitor traffic data rather the content of communications.[43] Unlike the Advocate General, the Court of Justice of the European Union doesn't discuss case law from Straatsburg. Likewise, the Court doesn't discuss the E-Privacy Directive, although the Court mentions it in the introduction of the judgment. A filter system like Sabam requested entails the use of deep packet inspection, a system that automatically inspects internet communication.[44] Such a system is difficult to reconcile with the right to secrecy of communications, as protected in the E-Privacy Directive:

> Member States shall ensure the confidentiality of communications and the related traffic data by means of a public communications network and publicly available electronic communications services, through national legislation. In particular, they shall prohibit listening, tapping, storage or other kinds of interception or surveillance of communications and the related traffic data by persons other than users, without the consent of the users concerned, except when legally authorised to do so in accordance with Article 15(1). (…)[45]

Web surfing and peer-to-peer traffic are 'communication' as defined in the E-Privacy Directive.[46] But it follows from *Promusicae* that member states are allowed to establish exceptions on the obligation to guarantee the confidentiality of communications, also to enable copyright enforcement.[47] However, member states have to respect fundamental rights and the proportionality principle.

---

[39] Article 29 Working Party 'Opinion 4/2007 on the concept of personal data' (WP 136) 20 June 2007.

[40] Since June 2012 the Dutch Telecommunications Act says that such profiles are presumed to be personal data (article 11.7a). The official Dutch text can be found on http://wetten.overheid.nl/BWBR0009950.

[41] Proposal for a regulation on the protection of individuals with regard to the processing of personal data and on the free movement of such data, COM(2012) 11 final.

[42] *Liberty v United Kingdom* App no 58243/00 at [56]; *S. and Marper v United Kingdom* App no 30562/04 (ECtHR 4 December 2009) at [104-105].

[43] *Copland v. United Kingdom* App no 62617/00 (ECtHR 3 April 2007) at [41-44]; *Malone v. United Kingdom* App no 8691/79 (ECtHR 2 August 1984) at [83-84].

[44] See for an overview of discussions about deep packet inspection for copyright enforcement in Europe and the United States: M. Mueller, A. Kuehn and S. M. Santoso, "Policing the Network: Using DPI for Copyright Enforcement" [2012] 9 Surveillance & Society 4, 348-364.

[45] Article 5(1) E-Privacy Directive.

[46] Article 2(d) E-Privacy Directive.

[47] *Promusicae v Telefonica de Espana* [2008] E.C.D.R. 10 at [53].

## 6. Conclusion

The Sabam judgments affirm that the right to intellectual property has obtained the status of a fundamental right in the European Union. But the Court also confirms that the right to intellectual property right is not absolute. It should be balanced with fundamental rights of others. Furthermore, a far-reaching filter obligation like Sabam requested is not allowed under the E-Commerce Directive. These insights are not surprising.

It is still possible that access providers or social network sites employ filter systems to enforce copyright. The judgments don't preclude national judges from imposing narrower filter obligations. Furthermore, hosting or access providers could by-pass judicial control and start filtering on the basis of a voluntary agreement with copyright holders. The European Commission recently encouraged internet service providers to battle copyright infringement together with rights holders. It is questionable whether the rights of internet users would be respected in such cases.

Furthermore, when balancing fundamental rights, the Court emphasises the freedom to conduct a business of the internet service providers. The freedom to conduct a business will put less weight in the scale if filter systems become cheaper or easier to implement. In that case, the balancing act between the fundamental rights of right owners, intermediaries and internet users may have a different outcome.

To conclude, the judgments in these particular cases are reasonable. But the Court could have given more attention to the fundamental rights of internet users. Important questions remain unanswered. What are the implications of filter systems for the right to freedom of information, to the confidentiality of communications and to respect for private life? Should deep packet inspection even be considered to assist a private organisation with copyright enforcement? Unfortunately, the judgements do not make it much clearer where the balance should lie between the rights of internet users, internet service providers and rights holders. The Court gave two reasonable decisions, but little is won for privacy and freedom of information.

\* \* \*